\documentclass[twocolumn,showpacs,amsmath,amssymb,pre,aps]{revtex4}   %%  4-pages !!
\usepackage{graphicx}% Include figure files
\usepackage{dcolumn}% Align table columns on decimal point
\usepackage{bm}% bold math
\usepackage{color}% color text

\renewcommand{\vec}[1]{\mathbf{#1}}

\newif\ifgraph

\graphtrue             %
\begin{document}
\title{
Non-Gaussian Normal Diffusion in a Fluctuating Corrugated Channel}

\author{Yunyun Li$^{1}$, Fabio Marchesoni$^{1,2}$, Debajyoti Debnath$^{3}$, and
Pulak K. Ghosh$^{3}$\footnote{E-mail: pulak.chem@presiuniv.ac.in (corresponding author)}}
 \affiliation{$^{1}$ Center for Phononics and Thermal Energy Science,
 School of Physics Science and Engineering, Tongji University, Shanghai 200092,
 People's Republic of China}
 \affiliation{$^{2}$ Dipartimento di Fisica, Universit\`{a} di Camerino, I-62032 Camerino, Italy}
 \affiliation{$^{3}$ Department of Chemistry,
Presidency University, Kolkata 700073, India}

\date{\today}

\begin{abstract}
A Brownian particle floating in a narrow corrugated (sinusoidal) channel
with fluctuating cross section exhibits non-Gaussian normal diffusion.
Its displacements are distributed according to a Gaussian law for
very short and asymptotically large observation times, whereas a robust
exponential distribution emerges for intermediate observation times of the
order of the channel fluctuation correlation time. For intermediate to large
observation times the particle undergoes normal diffusion with one and the same
effective diffusion constant.  These results are analytically interpreted
without having recourse to heuristic assumptions. Such a simple model thus
reproduces recent experimental and numerical observations obtained
by investigating complex biophysical systems.

\end{abstract}
\maketitle

\section{Introduction} \label{intro}

Recent observations \cite{Granick1,Granick2,Bhatta,Sung1,Sung2,Granick3} of
particle diffusion in fluctuating crowded environments  manifestly contradict
the common belief that normal diffusion is associated with a Gaussian
distribution of spatial displacements. Indeed, if for simplicity we restrict
ourselves to one dimensional (1D) geometries, the displacement, $\Delta
x(t)=x(t)-x(0)$, of a standard overdamped Brownian particle suspended in a
homogeneous Newtonian fluid \cite{Gardiner} (i) grows with time according to
the Einstein-Stokes law, $\langle \Delta x^2(t) \rangle = 2Dt$; (ii) is
distributed according to a rescaled Gaussian probability density function
(pdf), $p(\Delta x/\sqrt{t})$ with half-variance $D$. Under these
circumstances, the random variable $\Delta x(t)$ is said to undergo Gaussian
normal (or Fickian) diffusion.

There is no {\it a priori} reason why the diffusion of a tracer in a time
varying  inhomogeneous medium should be Fickian. In real biophysical systems,
with increasing the observation time the rescaled displacement distributions,
$p(\Delta x/\sqrt{t})$, often develop prominent exponential tails, whereas
the tracers start diffusing linearly in time. Such transient tails disappear
only for asymptotically large observation times (at times hardly accessible to real
experiments \cite{Granick1}), when finally the $\Delta x$ distributions turn
Gaussian, as predicted by the central limit theorem, without appreciably
changing the underlying diffusion mechanism. Persistent diffusive transients of this type
have been detected in diverse experimental setups
\cite{Granick1,Granick2,Bhatta,other1,other3,other4}. Extensive numerical
simulations confirmed the occurrence of this remarkable phenomenon in
crowded environments consisting of slowly diffusing or changing
microscopic constituents (filaments \cite{Granick1, Bhatta}, large hard
spheres \cite{Sung1,Sung2,Granick3}, clusters \cite{Kegel,Kob}, and other
heterogeneities \cite{Tong}).

The current interpretation of picture above postulates the existence of one
or more relaxation processes affecting the suspension medium or the confining
geometry, where the tagged particle diffuses \cite{Granick1}. As long as the
fixed time interval, $t$, over which $\Delta x(t)$ is measured is of the
order of the relaxation time constant(s), $\tau$, the particle displacement
can obey a non-Gaussian statistics. Through what mechanism, under
these conditions the particle's diffusion retains its normal
character, may vary from case to case. To this purpose, a popular paradigm revolves around the
heuristic notion of {\it diffusing diffusivity} \cite{Slater}, whereas the
environmental fluctuations are modeled by means of an {\it ad hoc} random
particle diffusion constant, $D(t)$. On assuming that $D(t)$ is an
Ornstein-Ulenbeck process with average $D$ and time constant $\tau$, the
distribution $p(\Delta x/\sqrt{t})$ changes from exponential for $t \ll \tau$
to Gaussian for $t\gg \tau$. In both time regimes, the displacement diffusion
is normal, with $\langle \Delta x^2(t) \rangle = 2Dt$ \cite{Slater}. This
phenomenological description, together with its more refined variations
\cite{Metzler2,Jain1,Jain2,Tyagi,Luo,Sokolov,Metzler}, may qualitatively
interpret a conspicuous body of diverse experimental observations, but sheds
little light on the underlying microscopic mechanisms.

In this paper we investigate both numerically and analytically the directed
diffusion of an overdamped Brownian particle in a narrow quasi-1D corrugated
channel \cite{chemphyschem,PNAS} of fluctuating width. Such a time variable
geometry is inspired to cell biology \cite{RMP_BM,Lipowski} and models the
key ingredient of the phenomenon under study, namely  slow environmental
fluctuations. The relevant stochastic model is detailed in Sec. \ref{model}.
The simulation results  of Sec. \ref{results} reproduce the essentials of non-Gaussian normal
diffusion \cite{Slater} without having recourse to the paradigm of diffusing diffusion: (1) The
distribution of the particle displacements along the channel is Gaussian for
observation times either much shorter (local diffusion) or much longer than
the correlation time of the channel fluctuation (channel diffusion), and
exponential for a rather wide interval of intermediate observation times; (2)
A normal diffusion law with the same channel diffusion constant extends from
intermediate to large observation times, thus implying a nontrivial
relationship between displacement pdf's. The compatibility
of normal diffusion with different displacement statistics is discussed in
Sec. \ref{Laplace}; (3) These effects are
robust as long as fluctuations randomly open and close the channel
constrictions. As remarked in the concluding Sec. \ref{conclusions},
the compartmentalization of particle's diffusion thus emerges as a
prerequisite of non-Gaussian normal diffusion.

\section{Fluctuating channel model} \label{model}

The dynamics of an overdamped (or massless) Brownian particle in a channel is
modeled by the Langevin equation ${\vec {\dot r}}(t)= \sqrt{D_0}~{\bm
\xi}(t)$, where ${\vec r}=(x,y)$ are the particle's coordinates and the
translational fluctuations ${\bm \xi}(t)=[\xi_x(t), \xi_y(t)]$ are zero-mean,
white Gaussian noises with autocorrelation functions $\langle \xi_i(t)
\xi_j(0)\rangle = 2\delta_{i,j} \delta (t)$, with $i,j = x,y$. The strength
of $\xi_i(t)$ is the free-particle diffusion constant, $D_0$, which is
typically proportional to the temperature of the suspension fluid. Without
loss of generality, we considered a two-dimensional (2D) sinusoidal channel
with axis oriented along $x$ and symmetrically confined transverse
coordinate, $|y |\leq w(x,t)$, where
\begin{equation}\label{w}
  w(x,t)=(y_L/2)[\varepsilon^2+(1-\varepsilon^2)\sin^2(\pi x/x_L)].
\end{equation}
Here, $y_L$ and $x_L$ are respectively the maximum width and the length of
the unit channel cell, and $\varepsilon^2 y_L$ is the fluctuating width of the pores
located at $x=0 ~~{\rm mod}(\pi)$. We assume for simplicity that
$\varepsilon(t)$ obeys the Ornstein-Uhlenbeck equation
\begin{equation}\label{epsilon}
  \dot \varepsilon=-(\varepsilon -\varepsilon_0)/\tau +\sqrt{D_\varepsilon /\tau^2}~\xi_{\varepsilon} (t),
\end{equation}
where the noise $\xi_{\varepsilon} (t)$  has the same statistics of, but is
uncorrelated with the thermal noises, ${\bm \xi}(t)$. Unless stated
otherwise, we set $\varepsilon_0=0$, so that the average pore width is $\langle
\varepsilon^2 \rangle y_L$ with $\langle \varepsilon^2 \rangle \doteq
\sigma^2_{\varepsilon}=D_{\varepsilon}/\tau$. In order to ignore hydrodynamic
effects \cite{PNAS}, we addressed the case of pointlike particles in highly
viscous suspension fluids. Accordingly. we assumed that, for small channel
fluctuations, the varying pressure exerted by the walls on the fluid does not
sensibly modulate the particle's diffusion constant, $D_0$, neither in space
nor in time. In practice, to modulate the effective width of the channel pores
without incurring this difficulty, one can simply apply a
tunable external gating potential \cite{PNAS}.

We numerically integrated the particle Langevin equation in the free space
inside the channel by means of a Milstein algorithm \cite{Kloeden}; we
imposed reflecting boundary conditions at the channel's walls, $y=\pm
w(x,t)$, and took stochastic averages over not fewer than 10$^5$
particle's trajectories with random initial conditions.

\begin{figure}[tp]
\centering \includegraphics[width=8.0cm]{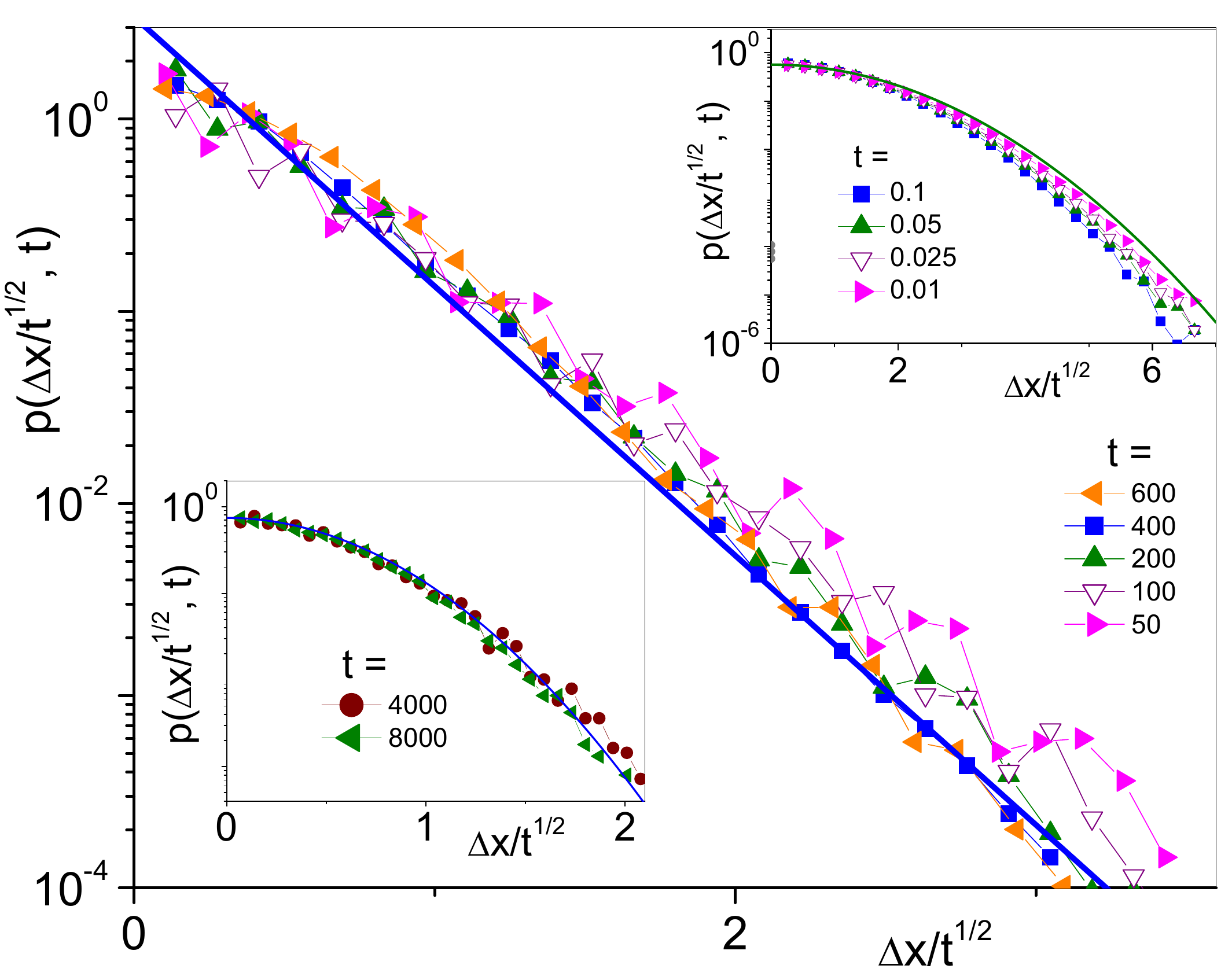}
\caption{Diffusion regimes in a fluctuating channel. The rescaled
displacement pdf's, $p(\Delta x/\sqrt{t})$, are computed for short, large
and intermediate observation times, $t$, respectively in the top-right inset,
bottom-left inset, and  main panel.
Simulation parameters are: $y_L=1$, $x_L=\pi$, $D_0=1$,
$D_{\varepsilon}=3$, $\varepsilon_0=0$, and $\tau=200$. The relevant Gaussian and Laplace
distributions are represented by solid curves. The half-variance of the Gaussian fitting
curves in the insets are $B=D_0=1$ (top-right) and $B=D=0.145$ (bottom-left);
the exponential decay constant is $\alpha =0.325$ (main panel). The parameters
$\alpha$ and $B$ are defined in the text. \label{F1}}
\end{figure}

\section{The case of opening-closing pores} \label{results}

The statistics of the particle displacement, $\Delta x(t)$, depends on the
observation time, $t$, as shown in Fig. \ref{F1}, where three different pdf
regimes are clearly distinguishable: two distinct Gaussian distributions,
$p(\Delta x/\sqrt{t}) = (4\pi B)^{-1/2}\exp({-\Delta x^2/4Bt})$, at very
short and large $t$ values (insets) and an exponential (or Laplace)
distribution, $p(\Delta x/\sqrt{t}) = (2\alpha)^{-1}\exp({-\Delta x/\alpha
\sqrt{t}})$, over an extended intermediate $t$ range. The short-$t$ Gaussian regime
describes the free Brownian diffusion inside a single channel cell, far from
the walls, which occurs for time intervals not larger than $\tau_L={\rm
min}\{x_L^2/8D_0, y_L^2/8D_0\}$. Under these circumstances, the fitting
parameter $B$ turned out to coincide with $D_0$,
 as expected \cite{Gardiner}.

For larger observation times the particle becomes sensitive to confinement
\cite{chemphyschem}. The escape from one cell into the adjacent ones requires
diffusing through narrow pores, a mechanism that takes relatively long
waiting times. On extending the approximate techniques of Ref. \cite{JCP137}
to the case of fluctuating pores, one estimates a characteristic mean-first
exit time (MFET)
\begin{equation}\label{tau0}
 \tau_0 = \frac{x_L^2}{8D_0}\frac{1}{\langle |\varepsilon| \rangle}=
 \frac{x_L^2}{8D_0}\sqrt{\frac{\pi \tau}{2D_{\varepsilon}}}.
\end{equation}
Here, $\tau_0$ is the time the particle takes to diffuse from inside a cell
up to the center of its left or right exit pore. Accordingly, the time
constant of the corresponding discrete intercell jumping process is $2\tau_0$
and the channel diffusion constant is thus well approximated by
\cite{PNAS,JCP137}
\begin{equation}\label{D}
D=x_L^2/4\tau_0.
\end{equation}
On the other hand, the correlation time of the fluctuating pore width,
$\varepsilon^2(t)$, is $2\tau$, see Eq. (\ref{epsilon}). In view of these
time scales, one is led to anticipate that the three distinct regimes of the
rescaled pdf illustrated in Fig. \ref{F1} must hold for observation times
$t\ll \tau_L$, $2\tau_0 \lesssim t \lesssim 2\tau$ and $t \gg 2\tau$,
respectively.
\begin{figure}[tp] \centering \includegraphics[width=8.5cm]{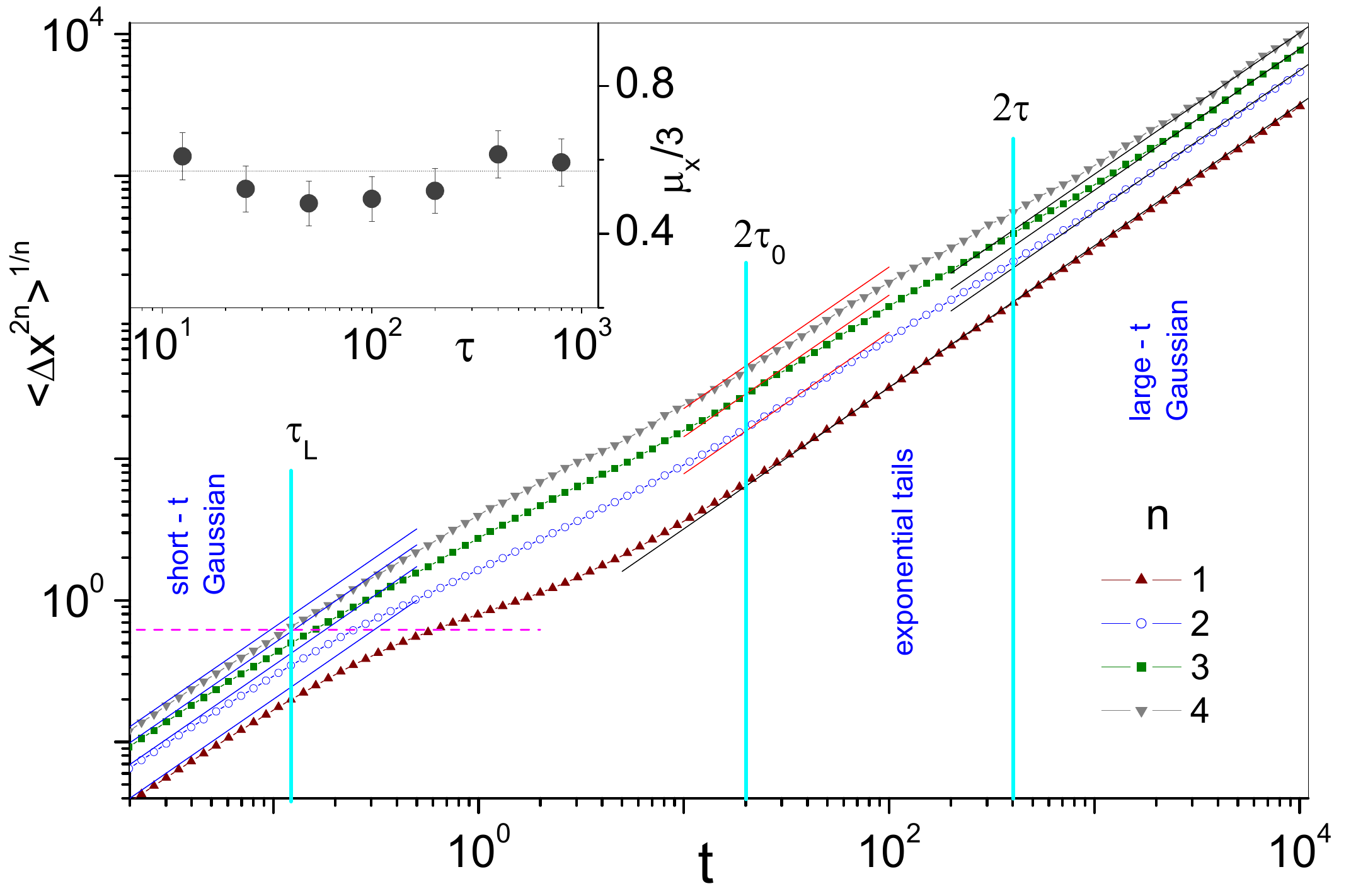}
\caption{Normal diffusion in a fluctuating channel: time dependence of
$\langle \Delta x^{2n}\rangle^{1/n}$ for different $n$. Simulation
parameters are: $y_L=1$, $x_L=\pi$, $D_0=1$, $\varepsilon_0=0$, $D_{\varepsilon}=3$, and
$\tau=200$.  The parallel straight lines represent the expected values for the
higher order moments ($n>1$) in terms of the fitted normal diffusion law
for $\langle \Delta x^{2}(t)\rangle$ in the Gaussian and exponential regimes (see text).
The vertical lines delimit the three pdf regimes of Fig. \ref{F1} and are positioned
respectively at $t=\tau_L, 2\tau_0$ and $2\tau$. Inset: ratio $\mu_x/3\simeq\alpha^2/B$ vs.
$\tau$ at large $t$. The numerical quantifier $\mu_x$ is defined in Sec. \ref{Laplace}
and was computed here for $t=2\tau$; for an ideal exponential transient
$\mu_x/3=\pi/2 -1$ (horizontal line).
 \label{F2}} \end{figure}

In the asymptotic regime, $t \gg 2\tau$, the particle diffusion along the
channel is normal (i.e., $\langle \Delta x^2 \rangle$ is a linear function of
$t$, Fig. \ref{F2}), which allows a direct numerical determination of the constant
$D$ of Eq. (\ref{D}). Moreover, the intercell jumping process yields a
Gaussian distribution of the discretized particle displacements
(Brownian random walker \cite{Gardiner}). This
suggests that asymptotically $B=D$, as discussed below.

The exponential transient, $2 \tau_0 \lesssim t \lesssim 2\tau$, thus bridges
two Gaussian limits, $t \to 0$ with $B=D_0$, and $t\to \infty$ with $B=D$.
Most remarkably, the same Laplace law (i.e., one decay constant $\alpha$)
fits all $\Delta x/\sqrt{t}$ distributions over about one decade of
observation times. The similarity with recent experimental observations is
apparent \cite{Granick1,Granick2,Bhatta,Sung1,Sung2,Granick3}. The short-$t$
Gaussian regime cannot be reproduced by the diffusing diffusion model of Ref.
\cite{Slater}, as there the intracell diffusion time, $\tau_L$, was
implicitly set to zero. We notice that the exponential transient interval can
be expanded by decreasing $\sigma^2_\epsilon$ while increasing $\tau$. In the
present case, however, this condition leads soon to extremely long simulation
runs.

A defining property of channel diffusion is featured in Fig. \ref{F2}: One
normal diffusion law with the same diffusion constant, $D$, fits all
numerical data for $\langle \Delta x^2 (t) \rangle$ with $t \gtrsim 2
\tau_0$, i.e., in correspondence with both the Laplace and the asymptotic Gaussian
distributions. To assess the Gaussian nature of the normal diffusion for $t \to 0$
and $t \to \infty$, we explicitly computed a few higher moments $\langle
\Delta x^{2n}\rangle$ with $n\geq 1$, also reported
 in Fig. \ref{F2}. We checked that in both limits $\langle \Delta
x^{2n}\rangle^{1/n}=[(2n-1)!!]^{1/n} \langle \Delta
 x^2\rangle$, as expected for Gaussian $\Delta x$ distributions.
 Instead, for the Laplace pdf's fitted
 in Fig. \ref{F1}, one would expect
 $\langle \Delta x^{2n}\rangle^{1/n}=[(2n!)/2^n]^{1/n} \langle \Delta
 x^2\rangle$. The agreement between these latter estimates and the actual diffusion data in
 the intermediate $t$ domain is qualitative good, only, which we attributed to
 the  deviations from the Laplace distributions, apparent
 at $\Delta x/\sqrt{t}\ll \alpha$.

 On the other hand, the Laplace and the large-$t$ Gaussian fitting pdf curves introduced
 above, yield the same channel diffusion constant, $D$,
only under the condition $\alpha^2=B$. However, the
 inset of Fig. \ref{F2} shows that, for the  parameters of Fig. \ref{F1},
 $0.8 < \alpha/\sqrt{B} < 0.9$. The ratio
 $\alpha/\sqrt{B}$ thus serves as a measure of the exponential character
 of the displacement statistics across the normal diffusion transient dominated by
channel fluctuations. An estimate of this ratio is obtained
 in Sec. \ref{Laplace}.

\begin{figure}[tp] \centering \includegraphics[width=8.0cm]{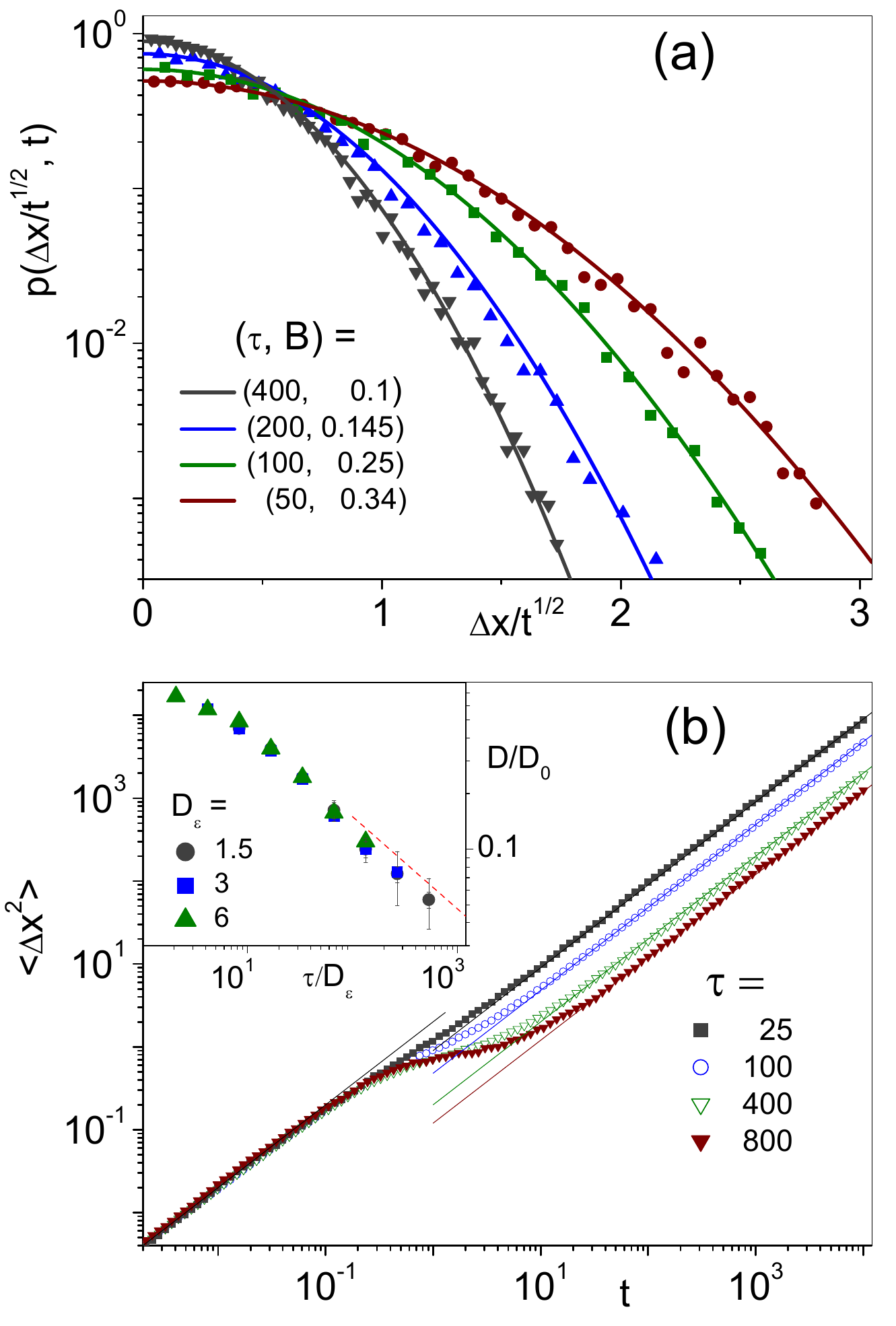}
\caption{Role of the fluctuation time correlation: (a) displacement pdf,
$p(\Delta x/\sqrt{t})$, and (b) diffusion, $\langle \Delta x^2\rangle$ vs.
$t$ for increasing $\tau$. Simulation parameters are: $y_L=1$, $x_L=\pi$,
$D_0=1$, $\varepsilon_0=0$, and $D_{\varepsilon}=3$. The straight lines on the r.h.s. of (b) are
the linear fits employed to extract the $D$ constant, whereas the solid lines
in (a) are the corresponding Gaussian curves, $p(\Delta x/\sqrt{t})$, defined
in the text, with $B=D$. For short $t$, all curves in (b) collapse on one
linear branch with diffusion constant $D_0$, also denoted by a straight line.
Inset: $D$ vs. $\tau$ for
different $D_\varepsilon$ (see legend). All other simulation parameters are as
in the main panel.
\label{F3}} \end{figure}

The role of the time scale $\tau$ is further illustrated in Fig. \ref{F3}.
As the MFET of Eq. (\ref{tau0}) increases like $\tau^{1/2}$, for
 large $\tau$ the diffusion curves $\langle \Delta x^2(t) \rangle$ develop a
 plateau in the range $\tau_L \ll t \ll \tau_0$. During this time interval,
 the diffusing particle ``fills up'' the 2D channel cell where it was initially
 injected, attaining a temporary maximum displacement $\langle \Delta
 x^2\rangle \simeq \bar x^2$, where $\bar x=\pi/4$ is the average half-width
 of the cell $w(x)$, Eq. (\ref{w}), for $\sigma_\varepsilon \to 0$. This
 diffusion plateau is represented in Fig. \ref{F2} by a horizontal line.

The numerical estimates of the channel diffusion constant, $D$, for $t
\gtrsim 2\tau_0$ and the fitting parameter $B$ of the Gaussian $\Delta x$
pdf for $t \gg 2\tau$ were anticipated to coincide. Simulation data. like those
reported in Fig. \ref{F3}, confirmed our expectations,
within the statistical error, for all
choices of the simulation parameters. Moreover, on combining Eqs. (\ref{tau0})
and (\ref{D}), one expects that $D/D_0 =2(2D_\varepsilon/\pi \tau)^{1/2}$,
also in fairly close agreement with the numerical data plotted in the inset
of Fig. \ref{F3}(b).

Finally, we notice that on lowering $\tau$ the slopes of the  normal
diffusion branches for $t \ll \tau_L$ and $t \gtrsim 2\tau_0$ tend to
coincide, that is $D$ (and $B$) tend to $D_0$. This is due to the fact that
the average pore width, $\langle \varepsilon^2\rangle y _L$, grows like
$D_\varepsilon/\tau$. As a result, for $\langle \varepsilon \rangle
\geqslant 1$ the effective channel bottlenecks are no longer located at $x=0
~{\rm mod} \{\pi\}$, but rather at $x=\pi/2 ~{\rm mod} \{\pi\}$, and have
fixed width, $y_L$. Accordingly, from Eq. (\ref{tau0}), $\tau_0=x_L^2/8D_0$
and the channel diffusion constant of Eq. (\ref{D}) is $D=D_0$.

\section{Exponential normal diffusion} \label{Laplace}

In Sec. \ref{results} we showed that on increasing the observation time larger
than twice the MFET, (i) the displacement distribution evolves from Laplacian
to Gaussian, and (ii) the same normal diffusion law holds for any $t$. A
simple argument, first introduced in Ref. \cite{Slater}, can be generalized
to support these conclusions.

Let us model a particle trajectory in the $x$ direction as the sum of random small
steps, $\Delta x_i$, taken at discretized times, $t_i=i\Delta t$, where $i=1, \dots N$
and $\Delta t=1$, for simplicity. The position of the particle at time $N$ is, therefore,
$x_N=\sum_{i=1}^{N} \Delta x_i$.
Accordingly,
\begin{equation} \label {sl1}
\langle x_N^2\rangle= \sum_{i=1}^{N}\langle \Delta x_i^2\rangle +
2\sum_{i \neq j}{}^\prime \langle \Delta x_i \Delta x_j\rangle,
\end{equation}
where $\sum_{i \neq j}^\prime $ stays for $\sum_{i=1}^{N-1}\sum_{j=1+1}^{N}$.
Contrary to the standard model of Brownian random walker \cite{Gardiner},
normal diffusion at time $N$ sets in under the generic condition that the
{\it step directions are uncorrelated}, $\langle \Delta x_i \Delta x_j
\rangle=0$, that is for mirror-symmetric distributions, $p(\Delta x_i)$, with
variances, $\langle \Delta x_i^2 \rangle$, possibly different, but of the
same order.

Following the authors of Ref. \cite{Slater}, one can further assume that during each unit time
step the particle's diffusion is normal  with time-dependent constant, $D_i$,
i.e., $$p(\Delta x_i) = (4\pi D_i)^{-1/2} \exp(-\Delta x_i^2/4D_i),$$ with
unspecified  $D_i$ distribution, $p(D_i)$. It follows immediately that
\begin{equation}\label{sl2}
\langle x_N^2\rangle = 2\langle D\rangle N,
\end{equation}
and

\begin{eqnarray}
\langle x_N^4\rangle - 3\langle x_N^2\rangle^2&=& 12(\langle D^2\rangle-\langle
D\rangle^2)N  \nonumber \\
&+&24\sum_{i \neq j}{}^\prime (\langle D_i D_j\rangle-\langle D_i\rangle \langle D_j\rangle),\label{sl4}
\end{eqnarray}
where $\langle D \rangle \equiv \langle D_i \rangle$ for the relevant choice of
$p(D_i)$.

Suppose now that two particle steps, $\Delta x_i$ and $\Delta x_j$ are
statistically uncorrelated, i.e., $\langle D_i D_j\rangle=\langle D_i
\rangle\langle D_j \rangle$, only for $|i-j|> \tau$. We then distinguish two
limiting cases,

(i) $N \gg \tau$, where
\begin{equation}\label{slG}
  \mu_x=\frac{\langle x_N^4\rangle - 3\langle x_N^2\rangle^2}{\langle x_N^2\rangle^2} = \frac{3 \mu_D}{N} \rightarrow 0,
\end{equation}
with $\mu_D= (\langle D^2\rangle-\langle D\rangle^2)/\langle D\rangle^2$. A
vanishing $\mu_{\rm x}$ hints at a Gaussian $x_N$  distribution as obtained
from numerical simulation;

(ii) $N \ll \tau$, where
\begin{equation}\label{slL}
  \mu_x=\frac{\langle x_N^4\rangle - 3\langle x_N^2\rangle^2}{\langle x_N^2\rangle^2}  \simeq 3 \mu_D.
\end{equation}
$\mu_{\rm x}=3$  would correspond to an exponential distribution of $x_N$; the
coefficient $\mu_D$ is thus a measure of the deviation of the actual $x_N$
distribution from the ideal Laplace distribution.

To apply the argument above to the model under study, the time step $\Delta
t$ has to be taken not shorter than $\tau_0$, i.e., the argument does not
hold for the intracell diffusion. The corresponding coefficient $\mu_D$ can
be estimated analytically by adapting the procedure of Ref. \cite{JCP137}
 to the case of a fluctuating channel, namely
\begin{equation}\label{slmD}
\mu_D \simeq \frac{\langle \varepsilon^2\rangle - \langle \varepsilon \rangle^2}{\langle\varepsilon\rangle^2} = \frac{\pi}{2} -1.
\end{equation}
On the other hand, on adopting the Laplace distribution $p(\Delta
x/\sqrt{t})$ for $x_N$ and the normal diffusion law, $\langle
\Delta x^2\rangle = 2Dt$, instead of Eq. (\ref{sl2}), that is, $\langle D
\rangle =D=B$, numerator and denominator of $\mu_{\rm x}$ can be calculated
explicitly to obtain
\begin{equation}\label{slmx}
  \mu_x =\frac{3\alpha^4}{B^2}.
\end{equation}
Finally, on comparing Eqs. (\ref{slL})-(\ref{slmx}), one  estimates
$\alpha/B^{1/2} \simeq 0.86$, in fairly close agreement with the numerical
data reported in the inset of Fig. 2.

\begin{figure}[tp] \centering \includegraphics[width=8.0cm]{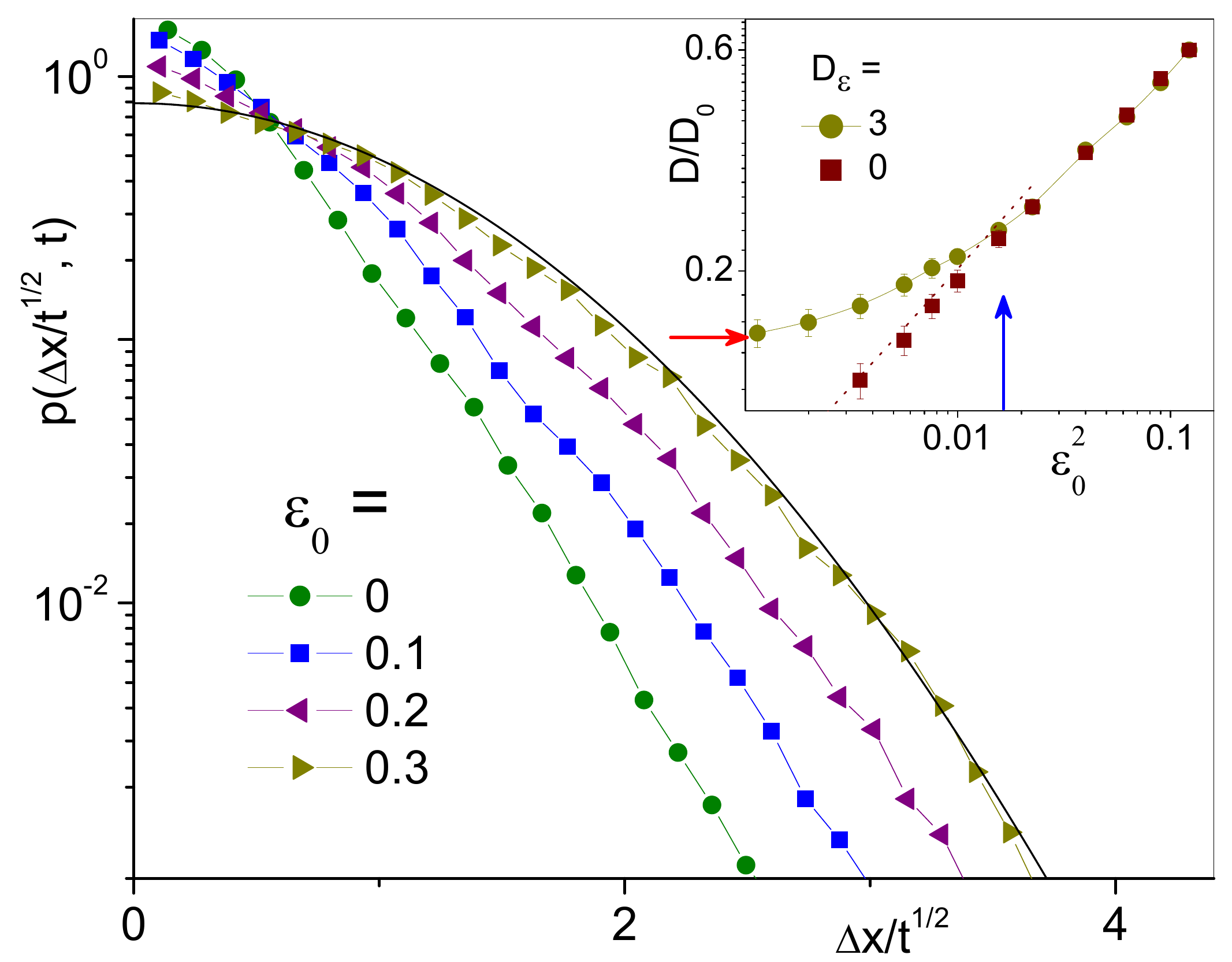}
\caption{Role of the fixed pore width $\varepsilon_0$: $p(\Delta x/\sqrt{t})$
at $t=2\tau$ for increasing $\varepsilon_0$ (see legend).
Other simulation parameters are: $y_L=1$, $x_L=\pi$, $D_0=1$,
$D_{\varepsilon}=3$,  and $\tau=200$. The solid curve represents a
normalized Gaussian distribution with $B=D$, $D$ being fitted from the normal
diffusion data (inset). Inset: channel diffusion constant,
$D$, as a function of $\varepsilon_0$ in the presence ($D_\varepsilon=3$) and absence
($D_\varepsilon=0$) of channel fluctuations. The dashed line is the analytical
prediction $D/D_0=2\varepsilon_0$ \cite{JCP137}; the values $\varepsilon_0^2=\sigma_\varepsilon^2$ and
$D/D_0$ at $\varepsilon_0=0$ are denoted by a vertical and a horizontal
arrow, respectively.
\label{F4}} \end{figure}

\section{Conclusions} \label{conclusions}

So far, by setting $\varepsilon_0=0$ in Eq. (\ref{epsilon}) we assumed that the
fluctuations of the channel cause the opening and closing of its pores. Of
course, in most cases the pores remain open at all times, with $\varepsilon_0>0$
and slightly fluctuating cross-section. Therefore, we investigated how the
non-Gaussian normal diffusion mechanism depends on $\varepsilon_0$. In the main
frame of Fig. \ref{F4} we plotted the rescaled displacement pdf's at
$t=2\tau$ for the simulation parameters of Fig. \ref{F1}, except that
$\varepsilon_0$ is increased from $0$ up to well above $\sigma_\varepsilon$. One
sees immediately that on widening the pores, the rescaled pdf's change from
exponential at $\varepsilon_0=0$, see Fig. \ref{F1}, to Gaussian with $B=D$, for
$\varepsilon_0 \gtrsim \sigma_\varepsilon$. The dependence of $D$ on $\varepsilon_0$
in the absence and presence of channel fluctuations is compared in the inset
of Fig. \ref{F4}. For $\varepsilon_0 \gtrsim \sigma_\varepsilon$ the diffusion
constant grows insensitive to the fluctuation strength, $D_\varepsilon$. The
obvious conclusion is that non-Gaussian normal diffusion only occurs when the
fluctuations of the channel walls are strong enough to actually open and
close the pores. Note that in the absence of channel fluctuations,
$D_\varepsilon=0$, our data for $D$ are well fitted by Eq. (\ref{D}), with
$\tau_0$ given by the first Eq. (\ref{F3}) upon replacing $\langle
\varepsilon\rangle$ with $\varepsilon_0$ \cite{JCP137}.

The results of Fig. \ref{F4} illustrate the importance of diffusion
compartmentalization during intermediate observation time intervals, $2 \tau_0
\lesssim t \lesssim 2\tau$. For $\varepsilon_0 > \sigma_\varepsilon$, the tagged
particle diffuses along the channel at all times, with only weakly correlated
open-pore crossings; hence a Gaussian displacement distribution. In sharp
contrast, for $\varepsilon_0 < \sigma_\varepsilon$, the pore crossings of the
trapped particle grow more and more time correlated; hence the exponential
tails of $p(\Delta x/\sqrt{t})$ discussed in Sec. \ref{Laplace}. This
description is consistent with the subordination mechanism advocated in Ref.
\cite{Sokolov}.

The microscopic model investigated in this paper, despite its simplicity,
was proven to reproduce most of the intriguing properties of the phenomenon
known as non-Gaussian normal diffusion. Such a phenomenon has emerged as
ubiquitous in soft matter physics, which suggests a number of promising
generalizations of the present model by incorporating additional sources of
randomness \cite{XX0}, for instance, by decorrelating channel's pore spacing
and fluctuations \cite{XX1} or exciting size and configurational fluctuations
of the diffusing particles \cite{XX2}, each on a suitably long time scale.
Their combined action is expected to make the conclusions of the present
study even more robust.

\section*{Acknowledgements}
Y.L. is supported by the NSF China under grant No. 11875201. P.K.G. is
supported by SERB Start-up Research Grant (Young Scientist) No.
YSS/2014/000853 and the UGC-BSR Start-Up Grant No. F.30-92/2015. D.D. thanks
CSIR, New Delhi, India, for support through a Junior Research Fellowship.

\end{document}